\documentclass[%
 aip,apl,
 amsmath,amssymb,
 reprint,
]{revtex4-2}

\usepackage{graphicx}
\usepackage{dcolumn}
\usepackage{bm}

\usepackage[utf8]{inputenc}
\usepackage[T1]{fontenc}

\begin{document}


\title[Revealing the Atomic Structure of NiO/Ga\textsubscript{2}O\textsubscript{3} Interfaces]{Revealing the Atomic Structure of NiO/Ga\textsubscript{2}O\textsubscript{3} Interfaces}

\author{Michelle A. Smeaton}
 \email{michelle.smeaton@nlr.gov}
 \affiliation{National Laboratory of the Rockies, Golden, CO, USA}
 
\author{Krishna Acharya}
 \affiliation{Metallurgical and Materials Engineering Department, Colorado School of Mines, Golden, CO, USA}

\author{Anna Sacchi}
\affiliation{National Laboratory of the Rockies, Golden, CO, USA}

\author{Renae N. Gannon}
\affiliation{National Laboratory of the Rockies, Golden, CO, USA}

\author{M. Brooks Tellekamp}
\affiliation{National Laboratory of the Rockies, Golden, CO, USA}

\author{Andriy Zakutayev}
\affiliation{National Laboratory of the Rockies, Golden, CO, USA}

\author{Vladan Stevanovic}
\affiliation{Metallurgical and Materials Engineering Department, Colorado School of Mines, Golden, CO, USA}

\author{Steven R. Spurgeon}
\affiliation{National Laboratory of the Rockies, Golden, CO, USA}
\affiliation{Metallurgical and Materials Engineering Department, Colorado School of Mines, Golden, CO, USA}
\affiliation{Renewable and Sustainable Energy Institute, University of Colorado Boulder, Boulder, CO, USA}

\date{\today}

\begin{abstract}
NiO/Ga\textsubscript{2}O\textsubscript{3} heterojunctions have garnered significant attention for use in power electronics due to the ultrawide bandgap and wafer-scale availability of Ga\textsubscript{2}O\textsubscript{3} and the controllable p-type doping of NiO. However, the structure of NiO/Ga\textsubscript{2}O\textsubscript{3} interfaces remains underexplored, largely due to the complexity of the junction between their dissimilar cubic and monoclinic crystal structures. Here we investigate the atomistic structure of the NiO/Ga\textsubscript{2}O\textsubscript{3} interface for (100), (\={2}01), and (001) oriented Ga\textsubscript{2}O\textsubscript{3} substrates using aberration-corrected scanning transmission electron microscopy (STEM) in combination with interface modeling and image simulations. We evaluate the abruptness and consistency of the interfaces and compare them to calculated interface models, proposing precise atomic structures and assessing potential structural variation arising from complexity of the monoclinic Ga\textsubscript{2}O\textsubscript{3} crystal structure. Our interface analysis supports increased focus on (100) oriented Ga\textsubscript{2}O\textsubscript{3} as a candidat for fabricating high quality, low defect density NiO/Ga\textsubscript{2}O\textsubscript{3} heterojunction devices. Importantly, we consider the effects of specimen thickness and 3D-to-2D projection during the STEM imaging process to differentiate such effects from real crystal variations. This work provides insight into the effect of substrate orientation on NiO film and interface quality, creating a pathway to improving heterojunction properties. It further highlights important considerations for interpretation of stability and interlayer phase formation in these interfaces, which is crucial for their integration into reliable and robust power electronic devices. 
\end{abstract}

\maketitle

Heterojunctions comprising NiO films grown on $\beta$-Ga\textsubscript{2}O\textsubscript{3} are commonly studied for use in power electronic devices. Ga\textsubscript{2}O\textsubscript{3} is a promising ultrawide bandgap (UWBG) material with a bandgap of 4.5-5.0 eV and a theoretical breakdown electric field of 8 MV/cm.\cite{green_-gallium_2022, chen_recent_2026} Though p-type doping of Ga\textsubscript{2}O\textsubscript{3} is difficult to achieve,\cite{lyons_survey_2018} p-n junctions can be fabricated using several wide bandgap oxides to form a heterojunction. NiO is a good candidate due to its favorable band alignment with Ga\textsubscript{2}O\textsubscript{3}, wide bandgap of 3.6-4.0 eV, and controllable p-type doping.\cite{lu_recent_2023} In recent years, the community has made significant progress toward understanding crystallinity, carrier transport, and band structure in NiO/Ga\textsubscript{2}O\textsubscript{3} heterojunctions. One facet of these heterojunctions still lacking detailed understanding is the atomic structure of the NiO/Ga\textsubscript{2}O\textsubscript{3} interface, which can strongly impact band alignment and associated electron and phonon transport.

Interface structure is critically important for achieving desired electronic properties and device stability, especially when integrated into vertical devices, which have many advantages in power electronics.\cite{wong_vertical_2020, wen_vertical_2024} Precise atomic positions, bonding environments, and variation therein impact local electronic structure, thus affecting transport properties. Such variation, as well as lattice defects, also give rise to trap states that affect heterojunction performance. Similarly, heat extraction from these high power junctions is strongly impacted by phonon scattering at the interface.

Another concern for heterojunction performance is formation of interlayer phases during operation. This is of particular importance for power electronic devices operated at high voltage and temperature. Recently, researchers have also identified a NiGa\textsubscript{2}O\textsubscript{4} spinel phase that formed during high temperature cycling of NiO films grown on Ga\textsubscript{2}O\textsubscript{3}(001).\cite{egbo_niga2o4_2024} The researchers hypothesized that this layer could be advantageous due to its p-type nature. However, effective use of this strategy in device design requires a detailed understanding of the interface structure and distribution of the NiGa\textsubscript{2}O\textsubscript{4}.

Most NiO/Ga\textsubscript{2}O\textsubscript{3} research thus far has focused on heterojunctions formed on (001)-oriented Ga\textsubscript{2}O\textsubscript{3} substrates due to the wide availability of bulk wafers. However, the atomic structure and lattice alignment at the interface is not well understood. Furthermore, systematic comparison of interface structure for NiO grown on different Ga\textsubscript{2}O\textsubscript{3} orientations is lacking, likely due to the complexity of the interfaces. While researchers have reported epitaxial NiO film orientations for Ga\textsubscript{2}O\textsubscript{3} (001), (100), (\={2}01), and (\={1}02) substrates and noted some disorder within a few nanometers of the interfaces,\cite{li_reproducible_2023, nakagomi_orientational_2016, nakagomi_crystal_2020, oshima_epitaxial_2023} precise interface structure analysis has been limited due to the challenge of sample preparation and image interpretation presented by this complex interface. 

In this work, we directly visualize and assess NiO/Ga\textsubscript{2}O\textsubscript{3} interface atomic structure for heterojunctions grown on (100) and (\={2}01) oriented Ga\textsubscript{2}O\textsubscript{3} in addition to the common (001) oriented Ga\textsubscript{2}O\textsubscript{3} using aberration-corrected scanning transmission electron microscopy (STEM). We analyze atomic-resolution STEM images of each interface and compare them with calculated interface models, providing new insight into atomic structure, abruptness, and variation. Importantly, we elucidate the effects of specimen thickness and projection effects on interpretation of interface structure in the STEM images, providing a pathway to better understand other complex heterointerfaces. 

\begin{figure*}
\includegraphics[width=6.5 in]{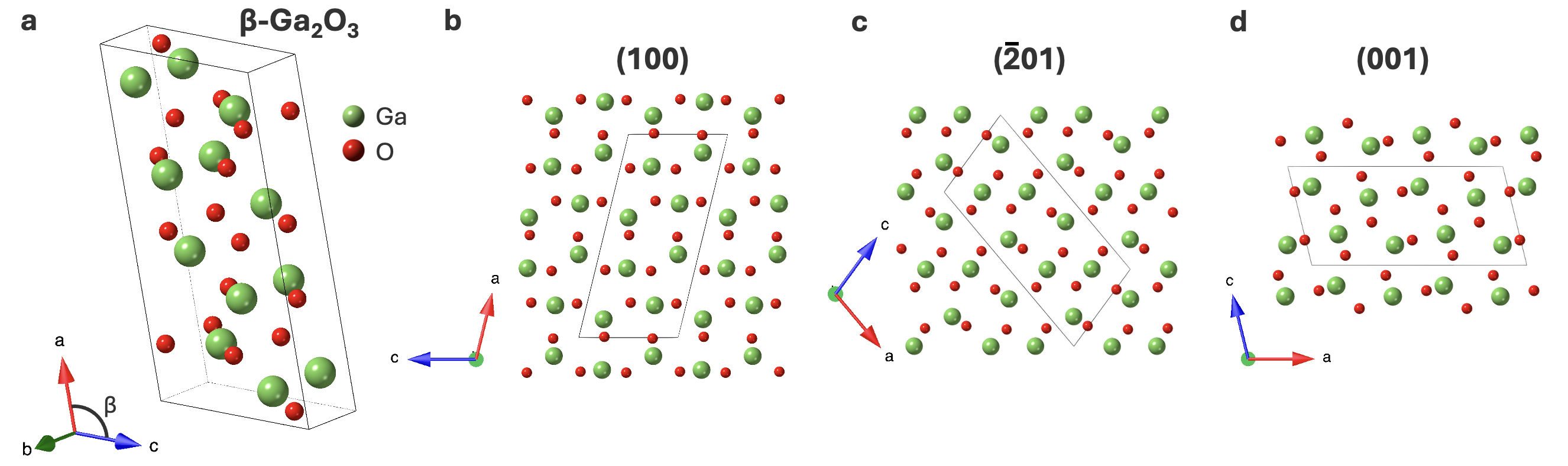}
    \caption{\label{fig:xtal_structure} Model of the $\beta$-Ga\textsubscript{2}O\textsubscript{3} crystal structure in (a) perspective view and viewed along the [010] crystallographic direction for the (b) (100), (c) (\={2}01), and (d) (001) substrate orientations.}
\end{figure*}

NiO films were grown by pulsed laser deposition (PLD) on commercially available $\beta$-Ga\textsubscript{2}O\textsubscript{3} substrates with (100), (\={2}01), and (001) nominal surface orientations. Prior to deposition, the substrates underwent organic solvent cleaning, \textit{i.e.}, sonication in acetone, methanol and isopropanol. The substrates were then In-bonded on a Si wafer and annealed at 350 °C in O\textsubscript{2} atmosphere for 10 minutes. During PLD growth, a NiO stoichiometric target was ablated with a pulsed KrF excimer UV laser ($\lambda$ = 248 nm) operated at 10 Hz frequency and 300 mJ energy. The growth was performed at a substrate temperature of 300 °C and O\textsubscript{2} partial pressure of 10 mTorr. The applied growth conditions yielded epitaxial NiO layers with a thickness of approximately 90 nm. 

While NiO has a cubic rock salt crystal structure, $\beta$-Ga\textsubscript{2}O\textsubscript{3} has a highly anisotropic structure with a monoclinic C2/m space group. It's pseudocubic oxygen sublattice, however, enables suitable epitaxial relationships with cubic crystal phases.\cite{oshima_epitaxial_2023} The structure is shown in Fig. \ref{fig:xtal_structure} in perspective view as well as in the three Ga\textsubscript{2}O\textsubscript{3} orientations examined here. All three oriented crystal models (Fig. \ref{fig:xtal_structure}(b-d)) are viewed along the [010] Ga\textsubscript{2}O\textsubscript{3} direction. STEM cross-sectional specimens were prepared for imaging along the same crystal direction. 

The epitaxial nature and crystal quality of all three films were confirmed using X-ray diffraction (XRD) (see Supplementary Material Fig. S1). These data also support our conclusions on NiO crystal orientation as discussed below. Following thin film characterization, cross-sectional lamella STEM specimens were prepared from each sample using a plasma focused ion beam (PFIB) lift out technique modified to both eliminate Ga alloying artifacts and minimize the effect of mechanical warping in very thin specimens. See Supplementary Material Section II and Figs. S2 and S3 for specimen preparation details. The final step of specimen preparation was to polish the lamella surfaces with a low energy (2 kV) Ar\textsuperscript{+} PFIB ion beam to remove remaining surface damage. 

High-angle annular dark-field (HAADF) STEM imaging was performed on a Thermo Fisher Scientific Spectra200 STEM operating at a 200 kV accelerating voltage with a convergence angle of 24.2 mrad. Moderate electron probe currents of 50 - 100 pA and dwell times of <1 \textmu s/px were utilized to obtain high signal-to-noise ratio images while ensuring negligible observable beam damage during data acquisition. Higher currents and dose rates were found to cause damage and evolution of the crystal lattice at the NiO/Ga\textsubscript{2}O\textsubscript{3} interface. 

\begin{figure*}
\includegraphics[width=7 in]{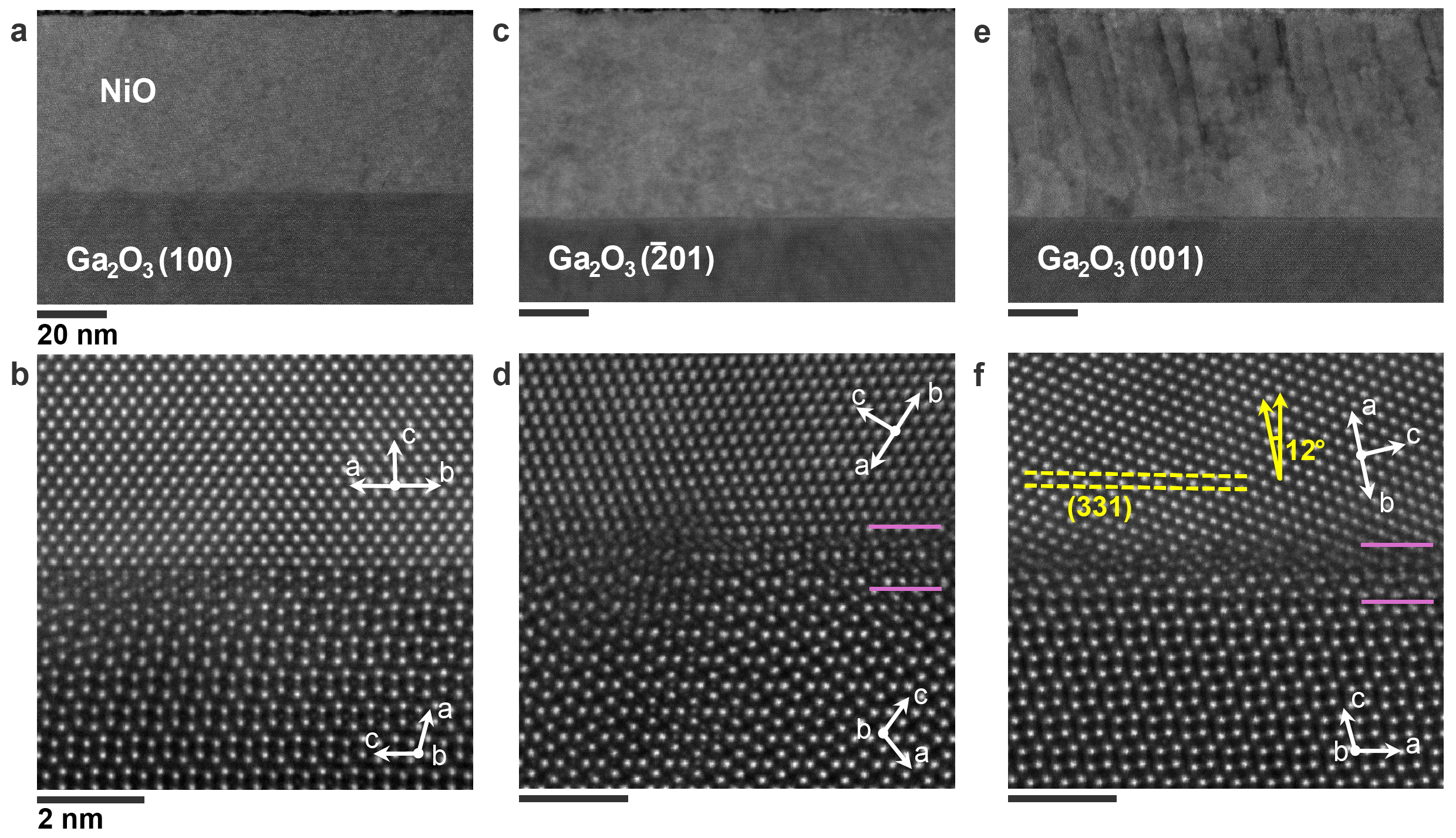}
\caption{\label{fig:overview} (a,c,e) HAADF-STEM images of the three NiO thin films capturing the full film thickness across over 100 nm. (b,d,f) Atomic-resolution images of the interface region of each sample. Images of the Ga\textsubscript{2}O\textsubscript{3}(100) oriented sample are shown in a,b; Ga\textsubscript{2}O\textsubscript{3}(\={2}01) in c,d; and Ga\textsubscript{2}O\textsubscript{3}(001) in e,f. Pink lines indicate layers of reconstructed lattice at the interface on (\={2}01) and (001) oriented Ga\textsubscript{2}O\textsubscript{3}. Scale bars in a,c,e are 20 nm, and scale bars in b,d,f are 2 nm.}
\end{figure*}

An overview image for each Ga\textsubscript{2}O\textsubscript{3} substrate orientation, showing the full NiO film thickness across >100 nm of the interface, is presented in Figs. \ref{fig:overview}(a,c,e). At this scale, all three interfaces look abrupt. The NiO films grown on (100) and (\={2}01) oriented Ga\textsubscript{2}O\textsubscript{3} (Figs. \ref{fig:overview}(a) and \ref{fig:overview}(c), respectively) are clearly single crystalline, while the film grown on (001) oriented Ga\textsubscript{2}O\textsubscript{3} (Fig. \ref{fig:overview}(e)) exhibits vertical extended defects approximately 20 nm above the interface. These defects suggest relaxation of strain that forms at the interface due to the epitaxial relationship with the (001) surface of Ga\textsubscript{2}O\textsubscript{3}. While investigation of these defects is outside the scope of this interface study, they may have important consequences for the function and stability of devices grown on this Ga\textsubscript{2}O\textsubscript{3} orientation.

Each interface is shown in more detail in Figs. \ref{fig:overview}(b,d,f). The NiO/Ga\textsubscript{2}O\textsubscript{3}(100) interface (Fig. \ref{fig:overview}(b)) is nearly atomically abrupt and qualitatively consistent across the 0.5 $\mu$m of total interface analyzed in this study (see Supplementary Material Fig. S4 for additional images). Even so, there are small regions of variation in the few atomic layers at the interface, for example on the left side of Fig. \ref{fig:overview}(b). The origin of these variations is explored below. The NiO/Ga\textsubscript{2}O\textsubscript{3}(\={2}01) and NiO/Ga\textsubscript{2}O\textsubscript{3}(001) interfaces (Figs. \ref{fig:overview}(d) and \ref{fig:overview}(f), respectively) are much more complex. They are notably less abrupt, each containing a handful of layers of reconstructed lattice denoted by the pink lines in Figs. \ref{fig:overview}(d) and \ref{fig:overview}(f). There is also more variation in the positions of atomic columns along these interfaces. The images presented here show the most common structure motifs observed for these samples. Additional images of each are presented in Supplementary Material Figs. S5 and S6. 

These atomic-resolution images enable identification of the NiO crystal orientation on each Ga\textsubscript{2}O\textsubscript{3} substrate. We observe a (001) growth orientation of NiO on Ga\textsubscript{2}O\textsubscript{3}(100), with an in-plane relationship of NiO [110] || Ga\textsubscript{2}O\textsubscript{3} [010]. For Ga\textsubscript{2}O\textsubscript{3}(\={2}01), the NiO growth orientation is (\={1}11), with NiO [110] || Ga\textsubscript{2}O\textsubscript{3} [010] in plane. These orientation relationships are consistent with our XRD measurements (Fig. S1 in the Supplementary Material) as well as previous reports based on XRD and electron diffraction.\cite{nakagomi_orientational_2016, nakagomi_crystal_2020, gong_band_2020} The NiO orientation on Ga\textsubscript{2}O\textsubscript{3}(001) is less clear, as also noted by others.\cite{nakagomi_crystal_2020, oshima_epitaxial_2023} As indicated in Fig. \ref{fig:overview}(f), the NiO [1\={1}0] crystallographic direction is tilted 12\textdegree{} from the growth direction. The orientation appears very close to (331) (yellow dashed lines in Fig. \ref{fig:overview}(f)). However, this is not quite right either. Based on HAADF-STEM images, we estimate the actual NiO orientation to be approximately (10 10 3). As in the other two samples, the in-plane orientation for Ga\textsubscript{2}O\textsubscript{3}(001) follows NiO [110] || Ga\textsubscript{2}O\textsubscript{3} [010]. 

Beyond extracting crystal orientation relationships between film and substrate, STEM enables examination of the precise atomic structure and alignment of the two lattices at the interfaces. This has become a common technique for investigating interface structure, defects, and charge transfer in interfaces between similar materials such as oxide perovskite layers.\cite{maclaren_aberration-corrected_2014, suyolcu_design_2020, nakagawa_why_2006, goodge_resolving_2023} However, the monoclinic Ga\textsubscript{2}O\textsubscript{3} to cubic NiO interface gives rise to much more complex epitaxial relationships, necessitating extreme care in sample preparation, imaging, and data interpretation. We further leverage interface structure calculations and STEM image simulations to compare with our experimental images, improving the confidence of our interpretation and providing insight into the energetic stability of the interfaces. The interface models were calculated using a previously described structure matching algorithm based on minimizing the Lennard Jones energy of the interface over a range of NiO surfaces for a given Ga\textsubscript{2}O\textsubscript{3} substrate,\cite{therrien_matching_2020} followed by density functional theory (DFT) structure relaxation. Details of the calculations are presented in Supplementary Material section IV.

\begin{figure*}
\includegraphics[width=6 in]{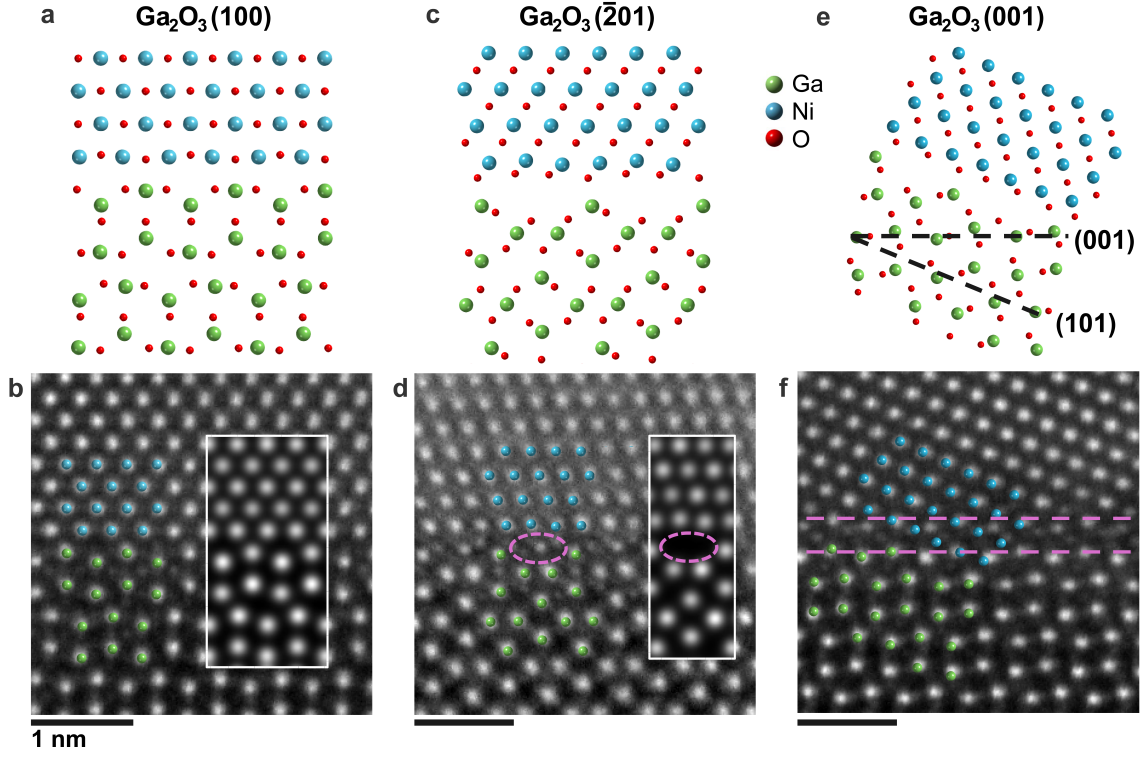}
\caption{\label{fig:model comparison} NiO/Ga\textsubscript{2}O\textsubscript{3} interface models compared with experimental HAADF-STEM images of the interface for (a,b) (100), (c,d) (\={2}01), and (e,f) (001) oriented Ga\textsubscript{2}O\textsubscript{3}. The calculated interface models are shown at in a, c, and e. b, d, and f show experimental HAADF-STEM images overlaid with the corresponding model (with oxygen atoms removed for clarity) and a simulated STEM image based on the model. For the Ga\textsubscript{2}O\textsubscript{3}(001) sample (e,f), a NiO/Ga\textsubscript{2}O\textsubscript{3}(101) model is shown instead, rotated to match the (101) plane in the image (f). Dashed pink ovals in d highlight an atomic column in the image that is not present in the NiO/Ga\textsubscript{2}O\textsubscript{3}(\={2}01) model, and dashed pink lines in f indicate the interface layers that are not well represented by the NiO/Ga\textsubscript{2}O\textsubscript{3}(101) model. Scale bars are 1 nm.}
\end{figure*}

Figure \ref{fig:model comparison}(a) presents the calculated model for NiO grown on Ga\textsubscript{2}O\textsubscript{3}(100). It is immediately clear that the calculation identified the same crystallographic orientation as was observed experimentally. Specifically, the interface is formed along the NiO (001) plane, and NiO [110] || Ga\textsubscript{2}O\textsubscript{3} [010]. Furthermore, the registry of atomic columns at the interface is in excellent agreement with the atomic columns in the image, as evidenced by the model overlaid on an experimental HAADF-STEM image in Fig. \ref{fig:model comparison}(b). For a more rigorous comparison, STEM images were simulated from the model using the multislice algorithm.\cite{goodman_numerical_1974, kirkland_simulation_1987, madsen_abtem_2021} The resulting simulated image is also overlaid on the experimental image in Fig. \ref{fig:model comparison}(a), again showing excellent agreement in contrast. Details of the multislice simulations are presented in Supplementary Material section V. There is some dim atomic column contrast between the bright Ga columns in the last couple Ga\textsubscript{2}O\textsubscript{3}(100) unit cells that does not appear in the simulated image. The contrast is likely due to a step edge in the Ga\textsubscript{2}O\textsubscript{3}(100) surface along the [010] direction and will be discussed in detail below. 

While the interface model for NiO/Ga\textsubscript{2}O\textsubscript{3}(100) appears to accurately represent the precise atomic interface structure in that sample, the calculations for Ga\textsubscript{2}O\textsubscript{3}(\={2}01) and Ga\textsubscript{2}O\textsubscript{3}(001) are somewhat more complicated. The model calculated for NiO/Ga\textsubscript{2}O\textsubscript{3}(\={2}01) is presented in Fig. \ref{fig:model comparison}(c). As was the case for Ga\textsubscript{2}O\textsubscript{3}(100), the calculation correctly identifies the observed growth and in-plane orientations of the NiO (NiO(\={1}11) with NiO [110] || Ga\textsubscript{2}O\textsubscript{3} [010] in plane). As shown by the atomic model overlaid on the experimental image in Fig. \ref{fig:model comparison}(d), the alignment of the NiO and Ga\textsubscript{2}O\textsubscript{3} crystal lattices is also well represented. In this case though, there appear to be additional atomic columns at the interface that are not present in the model (indicated by dashed pink ovals). We note that the brightest additional atomic column does not match the expected position of Ga in the Ga\textsubscript{2}O\textsubscript{3} lattice or known Ga interstitial sites (Figs. S8c and S8d in the Supplementary Material).\cite{johnson_unusual_2019, ingebrigtsen_impact_2018} This likely results from the uneven surface formed by the atoms in the Ga\textsubscript{2}O\textsubscript{3} (\={2}01) plane, which appears to be filled in by extra atoms as the film nucleates. 

Ga\textsubscript{2}O\textsubscript{3}(001) represents the most complex case, as has previously been noted.\cite{oshima_epitaxial_2023} The NiO growth plane has been reported as near (331),\cite{nakagomi_crystal_2020} but the true horizontal plane is higher order as indicated in Fig. \ref{fig:overview}(f). From STEM images, we estimate it to be the NiO (10 10 3) plane. Our interface structure calculations for Ga\textsubscript{2}O\textsubscript{3}(001) did not include high enough order planes to capture NiO (10 10 3) due to computational limitations. They did identify the (331) plane as being one of several relatively favorable NiO surfaces (see Supplementary Material Fig. S10(c)). However, the experimentally identified mistilt between the Ga\textsubscript{2}O\textsubscript{3} (001) and NiO (331) planes led us to investigate other potential epitaxial relationships for NiO on the Ga\textsubscript{2}O\textsubscript{3}(001) substrate. We identified such a relationship between the hexagonal close-packed planes on the oxygen sublattice of Ga\textsubscript{2}O\textsubscript{3} (101) and NiO (1\={1}1). The NiO(1\={1}1)/Ga\textsubscript{2}O\textsubscript{3}(101) relationship (Supplementary Material Fig. S1) as well as the lack of NiO(331)/Ga\textsubscript{2}O\textsubscript{3}(001) alignment were also verified by XRD. 

When the interface structure was calculated for the Ga\textsubscript{2}O\textsubscript{3} (101) surface, the model clearly identified (1\={1}1) as the NiO growth orientation, with NiO [110] || Ga\textsubscript{2}O\textsubscript{3} [010] in plane (Fig. \ref{fig:model comparison}(e)). When this model is overlaid on an experimental STEM image (Fig. \ref{fig:model comparison}(f)), the alignment of the Ga\textsubscript{2}O\textsubscript{3} and NiO lattices appears to match reasonably well. However, the couple atomic layers at the interface are not well captured, since the Ga\textsubscript{2}O\textsubscript{3} surface is different.  

\begin{figure*}[t!]
\includegraphics[width=\linewidth]{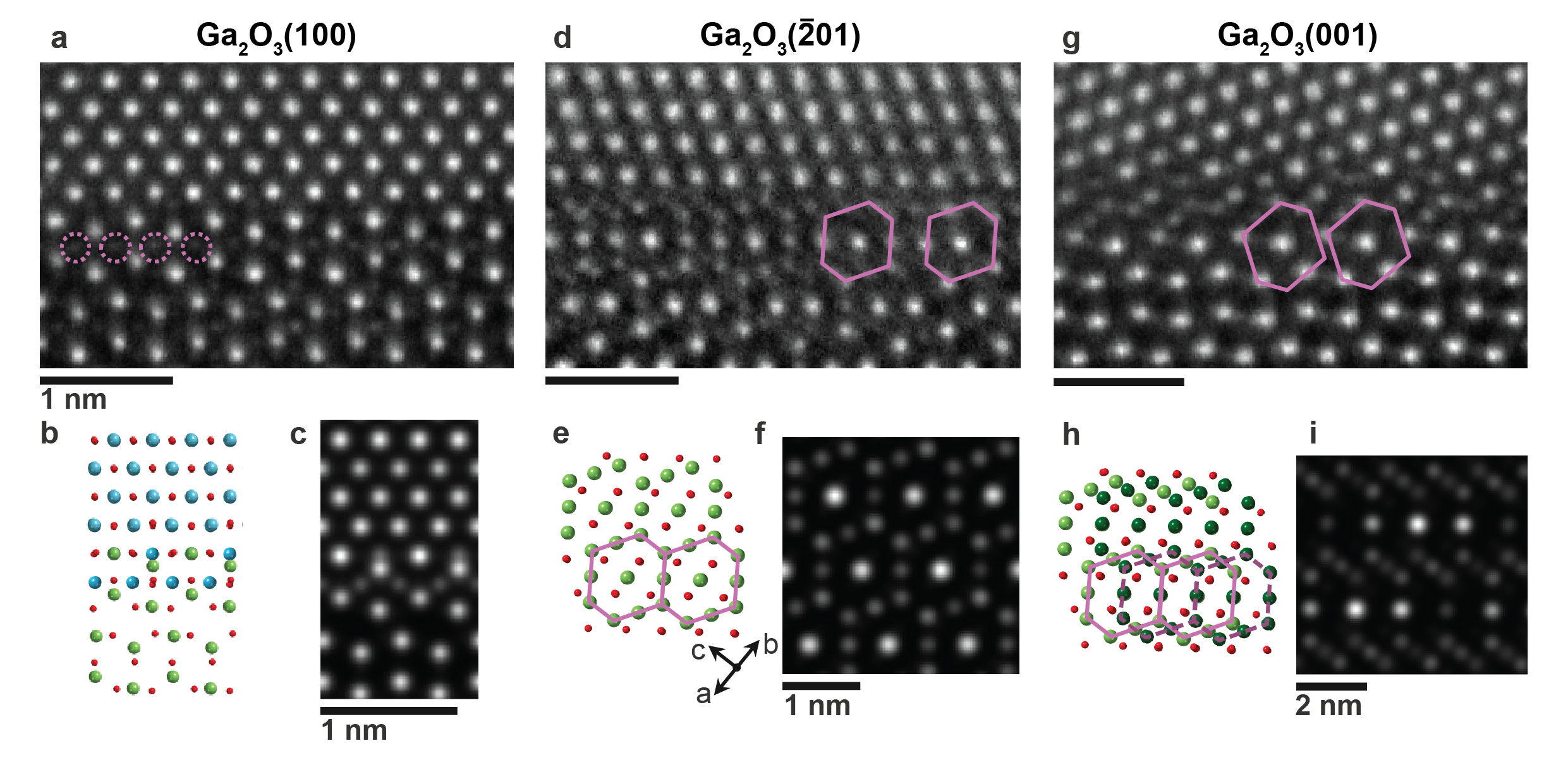}
\caption{\label{fig:artifacts} Importance of projection effects for interface interpretation. \textbf{(a)} HAADF-STEM image of NiO/Ga\textsubscript{2}O\textsubscript{3}(100) interface exhibiting weak atomic column intensity between Ga sites in the Ga\textsubscript{2}O\textsubscript{3} lattice. \textbf{(b-c)} Atomic model (b) and simulated STEM image (c) showing the calculated NiO/Ga\textsubscript{2}O\textsubscript{3} interface structure with a step edge along the viewing direction, resulting in a layer of Ni atoms between the top two layers of Ga sites. \textbf{(d)} HAADF-STEM imaging of NiO/Ga\textsubscript{2}O\textsubscript{3}(\={2}01) interface showing a hexagonal motif reminiscent of the spinel crystal structure. \textbf{(e-f)} Atomic model (e) and simulated STEM image (f) of the spinel ($\gamma$) polymorph of Ga\textsubscript{2}O\textsubscript{3} viewed along the [110] direction. Pink hexagons indicate the hexagon motif in both the spinel model (e) and experimental image (d), showing that the alignment of the motifs does not match. \textbf{(g)} HAADF-STEM image of a NiO/Ga\textsubscript{2}O\textsubscript{3}(001) interface exhibiting a different hexagonal motif, the alignment of which also does not match that in the $\gamma$-Ga\textsubscript{2}O\textsubscript{3} model in (e). \textbf{(h,i)} Model and simulated image of superimposed misaligned $\gamma$-Ga\textsubscript{2}O\textsubscript{3} layers, showing that the contrast also does not match the interfaces in (d) or (g).}
\end{figure*}

These data indicate that, during nucleation of the NiO film, the impinging atoms are influenced by the potential of multiple atomic layers of the Ga\textsubscript{2}O\textsubscript{3} surface. Thus, though we do not observe significant step edges exposing Ga\textsubscript{2}O\textsubscript{3}(101) surfaces, the NiO lattice forms with an epitaxial relationship between the tilted Ga\textsubscript{2}O\textsubscript{3}(101) and the NiO (1\={1}1). This relationship with a non-surface plane is further evidence that the Ga\textsubscript{2}O\textsubscript{3}(001) surface does not form an energetically favorable interface with NiO, leading to disorder and strain at the interface. Interfacial disorder and strain are evidenced both by variation in the atomic column motifs observed in STEM imaging and by variation in the alignment of the NiO lattice with the Ga\textsubscript{2}O\textsubscript{3}(101)/NiO(1\={1}1) interface model (see Supplementary Material Figs. S6 and S7).

Overall, this comparative interface analysis supports increased study of (100) oriented Ga\textsubscript{2}O\textsubscript{3} as a candidate for fabricating high quality, low defect density NiO/Ga\textsubscript{2}O\textsubscript{3} heterojunctions, though efforts are ongoing to improve the low homoepitaxial growth rate on this orientation.\cite{sasaki_mbe_2013, schewski_step-flow_2018, mazzolini_substrate-orientation_2020, jiang_single-crystalline_2025} NiO/Ga\textsubscript{2}O\textsubscript{3}(100) exhibits sharp crystallographic registry and limited defect density and structural variation as compared to NiO grown on (\={2}01) and (001) oriented Ga\textsubscript{2}O\textsubscript{3}. These results suggest better interface stability in Ga\textsubscript{2}O\textsubscript{3}(100) heterojunctions due to fewer nucleation sites for a NiGa\textsubscript{2}O\textsubscript{4} interlayer phase formation during high temperature device operation.

Analysis of complex heterointerfaces such as these hinges on meticulous STEM specimen preparation, image acquisition, and post-acquisition analysis. STEM is a projection technique, collapsing a three-dimensional structure into a two-dimensional image. Therefore, specimens must be extremely thin to ensure defects, disorder, and strain don't smear or blur out atomic columns, even when present in low density. Even so, in anything less than a perfectly abrupt and periodic interface or a unit-cell thick specimen, artifacts and misleading motifs can arise from this projection mechanism, leading to challenging interpretation. 

Figure \ref{fig:artifacts} presents examples of these effects identified in each of the NiO/Ga\textsubscript{2}O\textsubscript{3} interfaces examined here. In NiO grown on Ga\textsubscript{2}O\textsubscript{3}(100), a common variation in the generally abrupt interface is the presence of weak atomic column intensity between the Ga lattice sites in the top 1-2 unit cells of Ga\textsubscript{2}O\textsubscript{3} (Fig. \ref{fig:artifacts}(a)). This contrast initially appears like it could be due to interstitials in the Ga\textsubscript{2}O\textsubscript{3} lattice. Indeed, it matches a common Ga interstitial site, as shown in Figs. S8a and S8b of the Supplementary Material. However, closer inspection reveals that it is also explained by a step edge in the Ga\textsubscript{2}O\textsubscript{3} substrate along the viewing direction of the image, resulting in an overlap of the NiO and Ga\textsubscript{2}O\textsubscript{3} lattices. This is demonstrated in Figs. \ref{fig:artifacts}(b) and \ref{fig:artifacts}(c), which present an atomic model and a corresponding simulated STEM image, respectively. The model used to calculate the simulated image included a NiO layer shifted down by one unit cell that was 1/3 the total thickness of the model specimen. This led to the relatively low intensity of the extra Ni atomic columns in comparison to the rest of the Ni and Ga columns, which are very similar in intensity due to their close atomic numbers (28 and 31, respectively). Differentiating between Ni and Ga atoms in these sites would require atomically resolved electron energy loss spectroscopy (EELS) or energy dispersive X-ray spectroscopy (EDS) mapping, which is extremely challenging due to the beam sensitivity of these specimens. 

A common variation in the NiO/Ga\textsubscript{2}O\textsubscript{3}(\={2}01) interface is presented in Fig. \ref{fig:artifacts}(d). This region of the interface exhibits a hexagon motif that is reminiscent of the $\gamma$-Ga\textsubscript{2}O\textsubscript{3} polymorph,\cite{chang_-phase_2021} which has a defective spinel structure,\cite{mitome_rhombic_2013} or NiGa\textsubscript{2}O\textsubscript{4} phase, which has a very similar inverse spinel structure,\cite{greenwald_cation_1954} and has previously been reported in annealing studies of NiO/Ga\textsubscript{2}O\textsubscript{3} interfaces.\cite{egbo_niga2o4_2024} A model of this structure is shown in Fig. \ref{fig:artifacts}(e), viewed along the [110] direction and used to simulate a corresponding STEM image (Fig. \ref{fig:artifacts}(f)). The hexagon motifs are outlined in pink in the $\gamma$-Ga\textsubscript{2}O\textsubscript{3} model and the experimental STEM image, showing that their alignment is not the same. In the experimental image, adjacent hexagons have a space between, whereas they share a side (three Ga atoms) in the $\gamma$-Ga\textsubscript{2}O\textsubscript{3} model and simulated image. Thus, the contrast at the NiO/Ga\textsubscript{2}O\textsubscript{3} interface does not indicate formation of a $\gamma$-Ga\textsubscript{2}O\textsubscript{3} phase or NiGa\textsubscript{2}O\textsubscript{4} phase. Instead, the hexagon motif likely results from a similar projection effect as that depicted in Figs. \ref{fig:artifacts}(a-c). The exact atomic structure appears highly complex and outside the scope of the current analysis.

Similarly, Fig. \ref{fig:artifacts}(g) depicts a section of NiO/Ga\textsubscript{2}O\textsubscript{3}(001) interface which exhibits a different, partial hexagon motif. Here again, the spacing of the hexagons (shown in pink) does not match that in the $\gamma$-Ga\textsubscript{2}O\textsubscript{3} (Figs. \ref{fig:artifacts}(e) and \ref{fig:artifacts}(f)). We further compared the image contrast in the Ga\textsubscript{2}O\textsubscript{3} (\={2}01) and (001) cases (Figs. \ref{fig:artifacts}(d) and \ref{fig:artifacts}(g)) to a superposition of misaligned $\gamma$-Ga\textsubscript{2}O\textsubscript{3} layers, as shown in Figs. \ref{fig:artifacts}(h) and \ref{fig:artifacts}(i), which could reasonably form during recrystallization at the interface or during substrate polishing and preparation before NiO growth. Once again though, the model is clearly not a match to either experimental image, supporting our conclusion that a NiGa\textsubscript{2}O\textsubscript{4} interlayer phase has not formed in these as-deposited samples. Future work should apply this careful analysis to temperature cycled films to better understand potential evolution of an interlayer phase under device operating conditions. 

In summary, we have directly visualized and compared the atomic structure of NiO/Ga\textsubscript{2}O\textsubscript{3} interfaces grown on (100), (\={2}01), and (001) oriented Ga\textsubscript{2}O\textsubscript{3} substrates. We find a sharp interface in NiO/Ga\textsubscript{2}O\textsubscript{3}(100) with a clearly defined crystallographic registry, which we identified through interface structure calculations and confirmed by comparison with STEM images and image simulations. Ga\textsubscript{2}O\textsubscript{3} (\={2}01) and (001) form more complex interfaces with NiO. We identified the NiO orientation and epitaxial relationship on Ga\textsubscript{2}O\textsubscript{3} (\={2}01) but note that the ``corrugated'' surface of Ga\textsubscript{2}O\textsubscript{3} in this orientation results in added atomic positions not captured by our interface calculations. The NiO/Ga\textsubscript{2}O\textsubscript{3}(001) interface is the most complex, resulting in a near-(331) tilted NiO growth orientation approximated to the (10 10 3) crystallographic plane. We also found an epitaxial relationship between the Ga\textsubscript{2}O\textsubscript{3} (101) and NiO (111) planes that may drive initial NiO nucleation. Overall, these results support the adoption of (100) oriented Ga\textsubscript{2}O\textsubscript{3} as a candidate for fabricating high quality, low defect density NiO/Ga\textsubscript{2}O\textsubscript{3} heterojunctions. The sharp crystallographic registry and limited structural variation in these interfaces compared to those grown on Ga\textsubscript{2}O\textsubscript{3} (\={2}01) and (001) suggests better stability due to fewer nucleation sites for a NiGa\textsubscript{2}O\textsubscript{4} interlayer phase formation during high temperature device operation. Low homoepitaxial growth rates remain a challenge for efficient fabrication of power electronic devices.\cite{sasaki_mbe_2013, schewski_step-flow_2018} However, recent studies demonstrating improved Ga\textsubscript{2}O\textsubscript{3}(100) growth rates show it is still a promising route to high quality devices.\cite{mazzolini_substrate-orientation_2020, jiang_single-crystalline_2025} Regardless of Ga\textsubscript{2}O\textsubscript{3} substrate orientation, care must be taken in interpreting interface structure from STEM imaging, as specimen thickness and projection effects give rise to confusing atomic contrast. This point will be crucial for further studies of NiO/Ga\textsubscript{2}O\textsubscript{3} interface stability under device operation conditions. 

\begin{acknowledgments}
This work was supported as part of the A Center for Power Electronics Materials and Manufacturing Exploration (APEX) Energy Frontier Research Center funded by the U.S. DOE, Office of Science, Basic Energy Sciences. This work was authored in part by the National Laboratory of the Rockies (NLR) for the U.S. Department of Energy (DOE) under Contract No. DE-AC36-08GO28308. The views expressed in the article do not necessarily represent the views of the DOE or the U.S. Government
\end{acknowledgments}

\section*{Data Availability Statement}

The data that support the findings of this study are available from the corresponding author upon reasonable request.


\bibliography{Ga2O3, references}

\end{document}



\title{Supplementary Material: Revealing the Atomic Structure of NiO/Ga\textsubscript{2}O\textsubscript{3} Interfaces}

\author{Michelle A. Smeaton}
 \email{michelle.smeaton@nlr.gov}
 \affiliation{National Laboratory of the Rockies, Golden, CO, USA}
 
\author{Krishna Acharya}
 \affiliation{Metallurgical and Materials Engineering Department, Colorado School of Mines, Golden, CO, USA}

\author{Anna Sacchi}
\affiliation{National Laboratory of the Rockies, Golden, CO, USA}

\author{Renae N. Gannon}
\affiliation{National Laboratory of the Rockies, Golden, CO, USA}

\author{M. Brooks Tellekamp}
\affiliation{National Laboratory of the Rockies, Golden, CO, USA}

\author{Andriy Zakutayev}
\affiliation{National Laboratory of the Rockies, Golden, CO, USA}

\author{Vladan Stevanovic}
\affiliation{Metallurgical and Materials Engineering Department, Colorado School of Mines, Golden, CO, USA}

\author{Steven R. Spurgeon}
\affiliation{National Laboratory of the Rockies, Golden, CO, USA}
\affiliation{Metallurgical and Materials Engineering Department, Colorado School of Mines, Golden, CO, USA}
\affiliation{Renewable and Sustainable Energy Institute, University of Colorado Boulder, Boulder, CO, USA}

\date{\today}

\maketitle

\renewcommand{\thefigure}{S\arabic{figure}}

\section{Thin film characterization}

\begin{figure*}
\includegraphics[width=6.5 in]{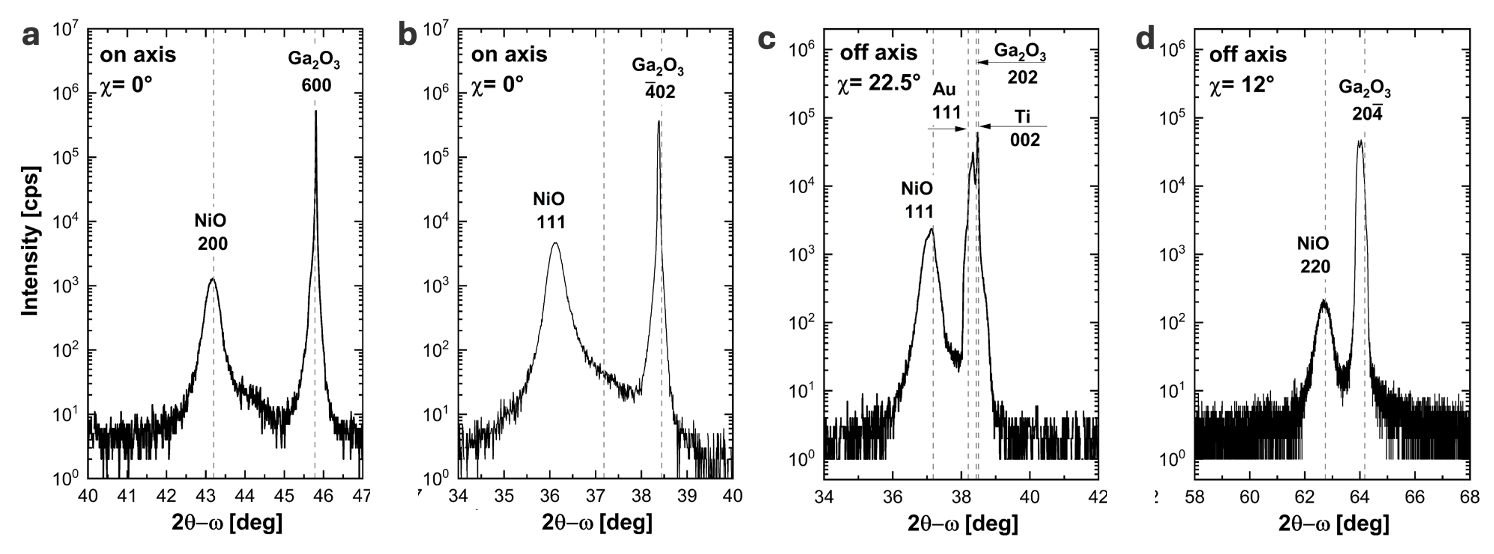}
\caption{\label{fig:XRD} \textbf{(a)} and \textbf{(b)} symmetric on-axis 2$\theta$-$\omega$ scans of NiO grown on (100) and (\={2}01) out-of-plane oriented-Ga\textsubscript{2}O\textsubscript{3}, respectively. \textbf{(c)} and \textbf{(d)} symmetric 2$\theta$-$\omega$ scans of off-axis (101) and (10\={2}) Ga\textsubscript{2}O\textsubscript{3} planes, for a NiO film grown on a (001) oriented substrate. The scans allow us to probe (111) and (110) NiO planes. The broadening of the substrate peaks in (c) and (d) is due to the presence of the Ti/Au contacts deposited on the sample. The vertical dashed lines identify the nominal position expected for each peak labeled.}
\end{figure*}

The crystal structures and epitaxial relationships for NiO epitaxial films grown on the different  Ga$_{2}$O$_{3}$ out of plane (OOP) orientations ((100), (\={2}01), (001)) were probed with X-ray diffraction. A Rigaku Smartlab diffractometer with a Cu K$\alpha$ source and a Ge (220) monochromator was used. On-axis, \textit{i.e.}, at $\chi$ = 0°, symmetric 2$\theta$-$\omega$ scans were used to probe NiO epitaxial layers grown on (100) and (\={2}01) oriented substrates. In Fig. \ref{fig:XRD} (a) and (b), we report that the epitaxial relationships for NiO on these two Ga$_{2}$O$_{3}$ orientations are as follows: NiO (100) || Ga$_{2}$O$_{3}$ (100) and NiO (111) || Ga$_{2}$O$_{3}$ (\={2}01). Film and substrate peaks are labeled and their nominal positions are highlighted by vertical gray dashed lines. In Fig. \ref{fig:XRD} (b), the shift observed for the NiO (111) diffraction peak, with respect to the nominal position, is related to the O\textsubscript{2} partial pressure during film growth. Although the nominal oxygen pressure (O\textsubscript{2} = 10 mTorr) was kept unchanged for all three NiO samples, variations in the growth chamber conditions may have influenced the crystalline quality and precise stoichiometry of the epitaxial layers. 

Characterization of the NiO film grown on (001)-oriented Ga$_{2}$O$_{3}$ required more elaborated measurements. The difficulty to detect an OOP oriented vector for NiO, aligned with the normal of the substrate, limited the XRD characterization to off-axis reflections, \textit{i.e.}, at $\chi$ $\neq$ 0°. Figure \ref{fig:XRD} (c) and (d) shows Ga$_{2}$O$_{3}$ (202) and (20\={4}) diffraction peaks probed at $\chi$ =  22.5° and 12°, respectively, where NiO (111) and (220) are detected. The relationship between NiO (111) and Ga$_{2}$O$_{3}$ (202) (equivalent to (101)) is discussed in detail in the main text. 

\section{TEM lamella preparation}

Electron-transparent lamellae of NiO/Ga$_{2}$O$_{3}$ were prepared with a Thermo Fisher Scientific Helios 5 Laser Hydra, a multiple ion species Plasma Focused Ion Beam (PFIB) system. To maintain the intrinsic structural and chemical properties of the NiO/Ga$_{2}$O$_{3}$ interfaces, inert plasma (Xe and Ar) species were selected for specimen preparation. Traditional Ga liquid metal ion sources (LMIS) frequently induce artifacts through Ga implantation into the sample surface. In this system, implanted would be indistinguishable from the Ga within the Ga$_{2}$O$_{3}$ substrates, complicating local chemical and structural analysis of the substrate. Furthermore, implanted Ga is highly reactive and has a strong propensity to alloy with transition metals, risking the formation of secondary phases (such as Ni-Ga alloys). By utilizing inert species PFIB, chemical contamination and beam-induced alloying effects were effectively eliminated.

\begin{figure*}
\includegraphics[width=6.3in]{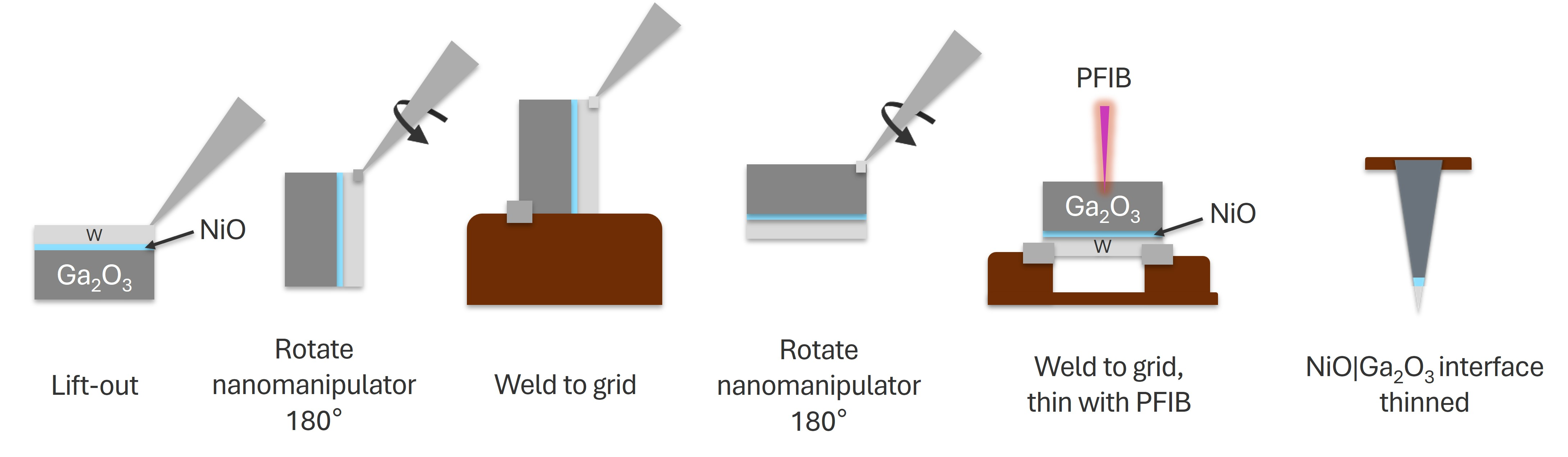}
\caption{Depiction of inverted lamella preparation process on a Thermo Fisher Scientific Helios 5 Laser Hydra equipped with an EasyLift EX nanomanipulator with motorized rotation}
\label{fig:InvertedPrep}
\end{figure*}

Despite the advantages of plasma focused ion beam (PFIB) preparation, achieving ultra-thin S/TEM lamellae ($<$ 30 nm) presents challenges when compared to conventional Ga FIB workflows. Plasma FIBs generally exhibit larger spot sizes and broader beam tails, particularly at lower accelerating voltages when compared to Ga FIBs. This can make precise end-pointing and final thickness control significantly more difficult, decreasing success rates for ultra-thin lamellae. To compensate for these challenges, we used an inverted preparation geometry (backside thinning) approach to mitigate curtaining effects and reproducibly achieve $<$ 50 nm lamellae.

Inverted lamella preparation on our system involves a combination of lift-out and rotation steps, depicted in Fig.~\ref{fig:InvertedPrep}. To protect the NiO film from ion beam-induced damage and implantation during the milling process, protective capping layers were deposited \textit{in situ}. Initially, a 200 nm thick Pt layer was deposited via electron beam-induced deposition. This was immediately followed by the deposition of a 3 $\mu$m thick W layer using Xe$^{+}$ PFIB at 12 kV. Bulk trenches were prepared with Xe$^{+}$ PFIB at 30 kV. Thinning of the lamella was then performed from the substrate side. 

\begin{figure}
    \centering
    \includegraphics[width=1\linewidth]{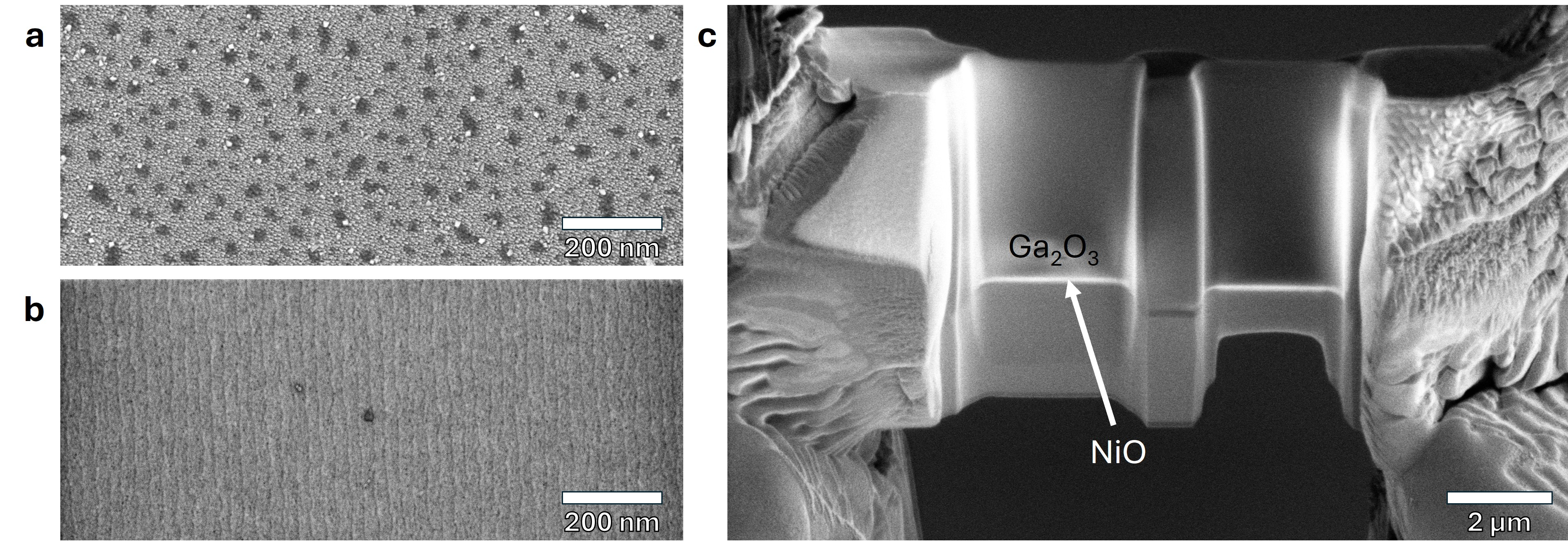}
    \caption{Example Secondary Electron SEM images of NiO/Ga$_{2}$O$_{3}$ samples. Top-down images of NiO on (001) and (\={2}01) Ga$_{2}$O$_{3}$  are shown in (a) and (b), respectively. Two thinned windows on an inverted lamella are shown in (c).}
    \label{fig:Lamella}
\end{figure}

The majority of the backside thinning was conducted using the Xe$^{+}$ PFIB at an accelerating voltage of 16 kV and currents ranging from 210 to 20 pA. High overtilt angles ($\pm$2.2--4.5$^\circ$) were used to create a wedge at the bottom of the chunk to target thinning only of the NiO/Ga$_{2}$O$_{3}$ interface. Maintaining a 16 kV accelerating voltage during the bulk thinning and keeping the bulk substrate thicker were critical to minimize damage and mechanical warping of the lamella. Additionally, the width of thinned windows was 3 $\mu$m at maximum, which minimized catastrophic lamella warping that would result in severe bowing of samples $<$ 100 nm. Then, brief polishes at 8 kV and 5 kV ($\pm$3--4$^\circ$) and currents ranging from 21--60 pA using Xe$^{+}$ PFIB were done to remove sidewall damage from 16 kV steps. For the final step, Ar$^{+}$ PFIB at 2 kV 25 pA  ($\pm$7$^\circ$) was used to remove remaining sidewall damage from primary Xe$^{+}$ PFIB steps. Using this conservative and carefully monitored process enabled reproducible preparation of ultra-thin lamellae while preserving intrinsic properties of the NiO/Ga$_{2}$O$_{3}$ interfaces. Example Scanning Electron Microscope (SEM) images of sample surfaces prior to lamella preparation are shown in Fig.~\ref{fig:Lamella} (a) and (b), and a SEM image of an example of a NiO/Ga$_{2}$O$_{3}$ inverted lamella with two < 50 nm thick windows is shown in (c).

\section{Additional STEM imaging}

\begin{figure*}
    \centering
    \includegraphics[width=\linewidth]{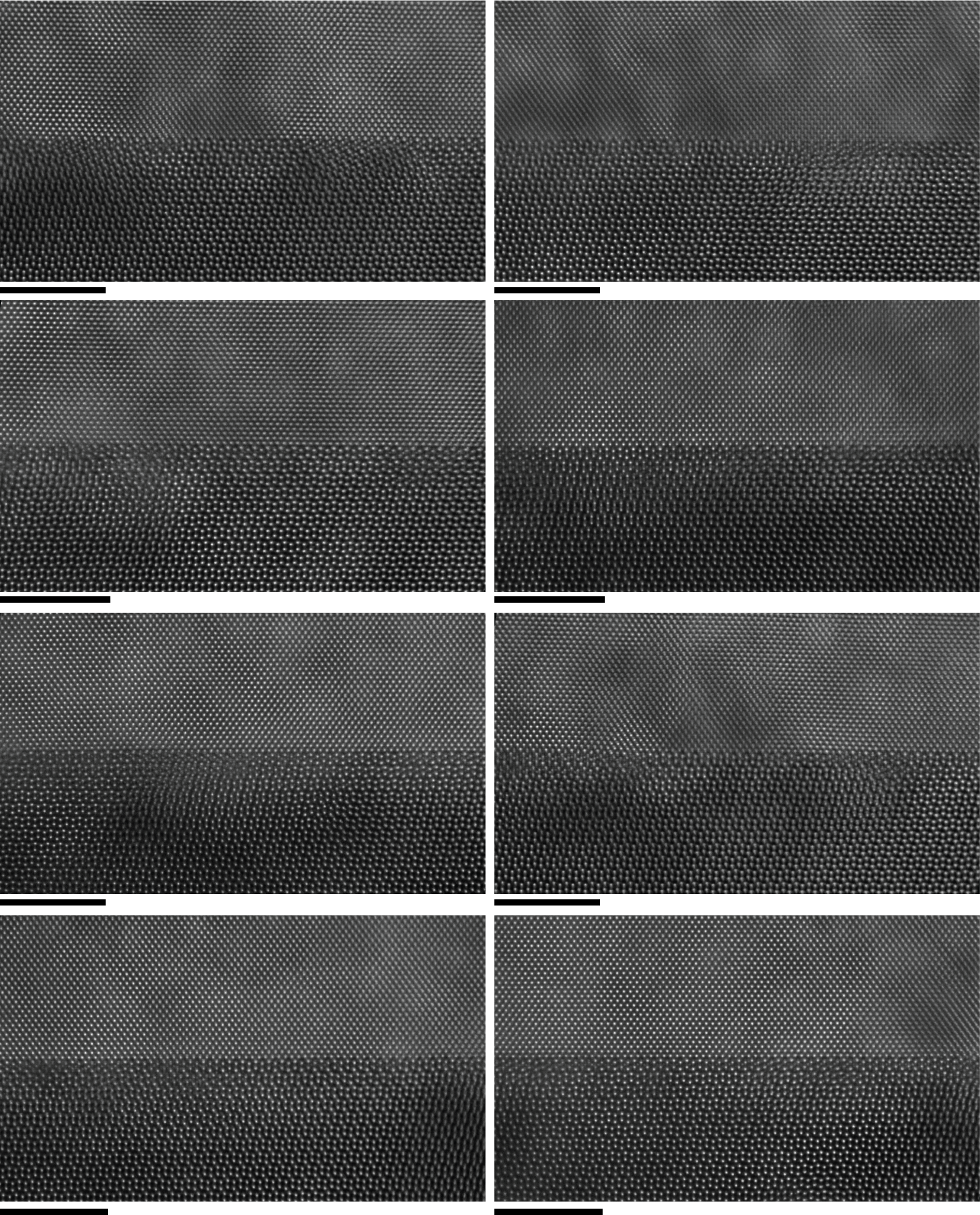}
    \caption{Additional representative images of the NiO/Ga$_{2}$O$_{3}$(100) interface, showing $\sim$ 175 nm of the $\sim$ 500 nm analyzed. Scale bars are 5 nm.}
    \label{SI_100}
\end{figure*}

\begin{figure*}
    \centering
    \includegraphics[width=6.5in]{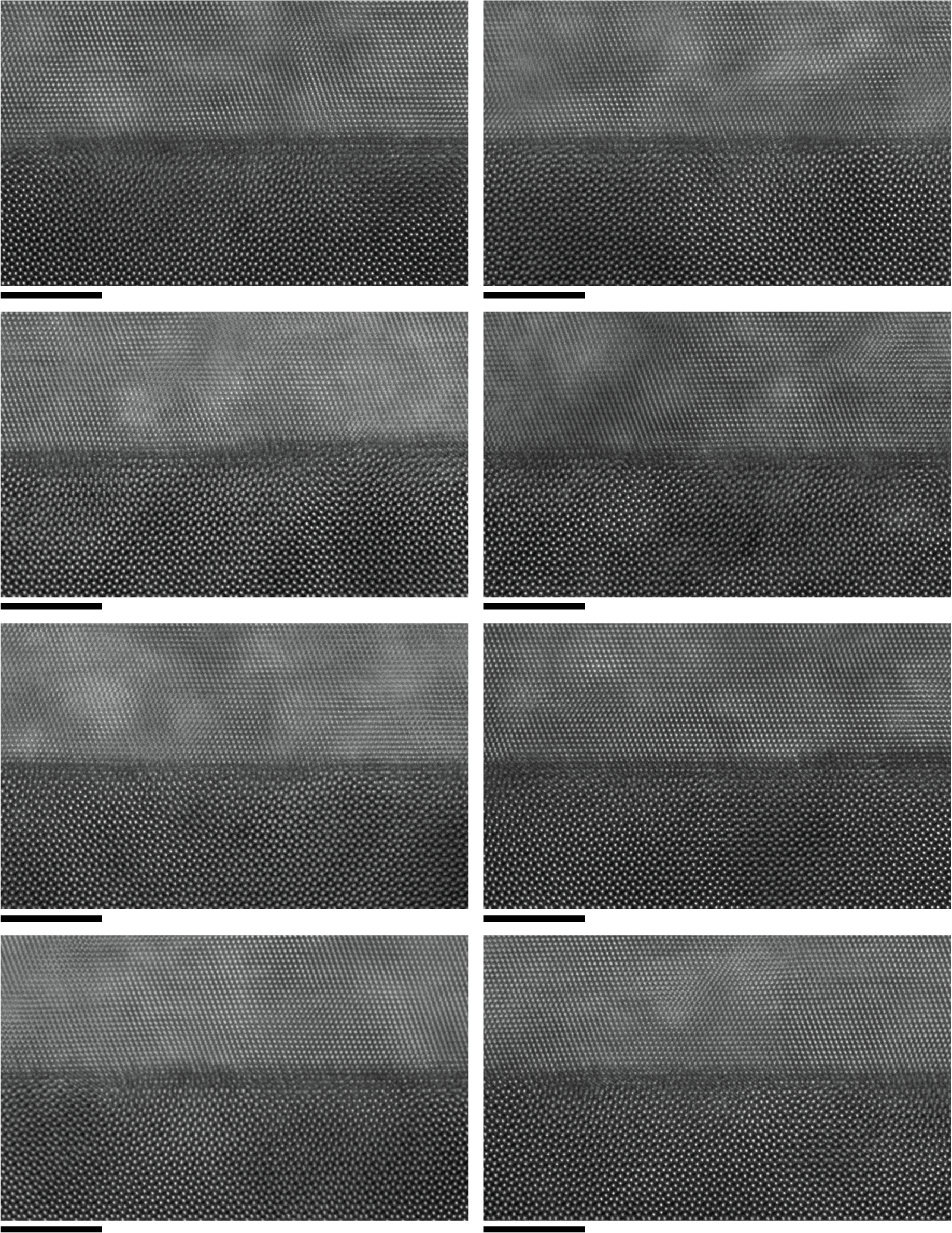}
    \caption{Additional representative images of the NiO/Ga$_{2}$O$_{3}$(\={2}01) interface, showing $\sim$ 175 nm of the $\sim$ 500 nm analyzed. Scale bars are 5 nm.}
    \label{SI_-201}
\end{figure*}

\begin{figure*}
    \centering
    \includegraphics[width=6.5in]{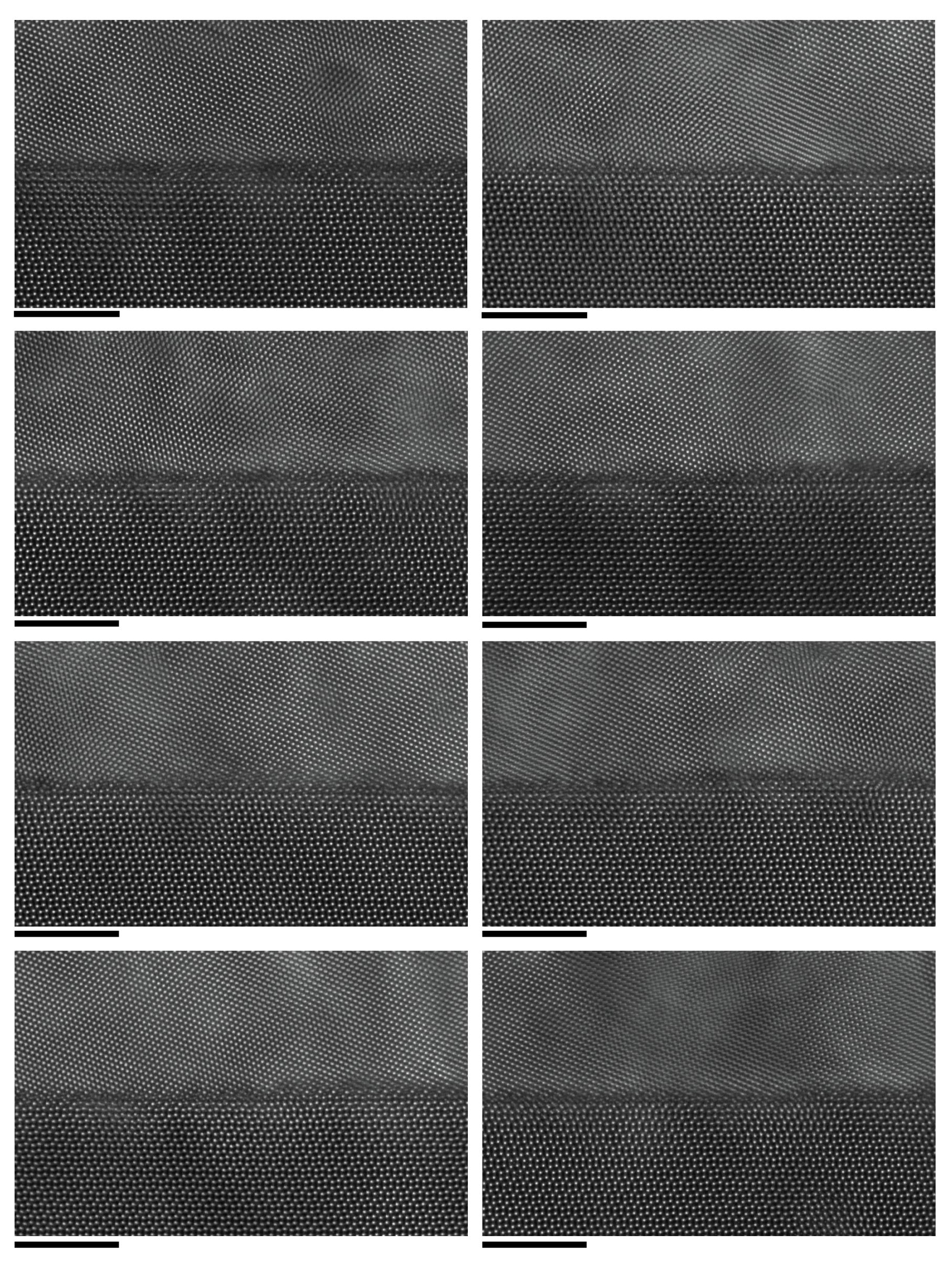}
    \caption{Additional representative images of the NiO/Ga$_{2}$O$_{3}$(001) interface, showing $\sim$ 175 nm of the $\sim$ 500 nm analyzed. Scale bars are 5 nm.}
    \label{SI_001}
\end{figure*}

\begin{figure*}
    \centering
    \includegraphics[width=6.5in]{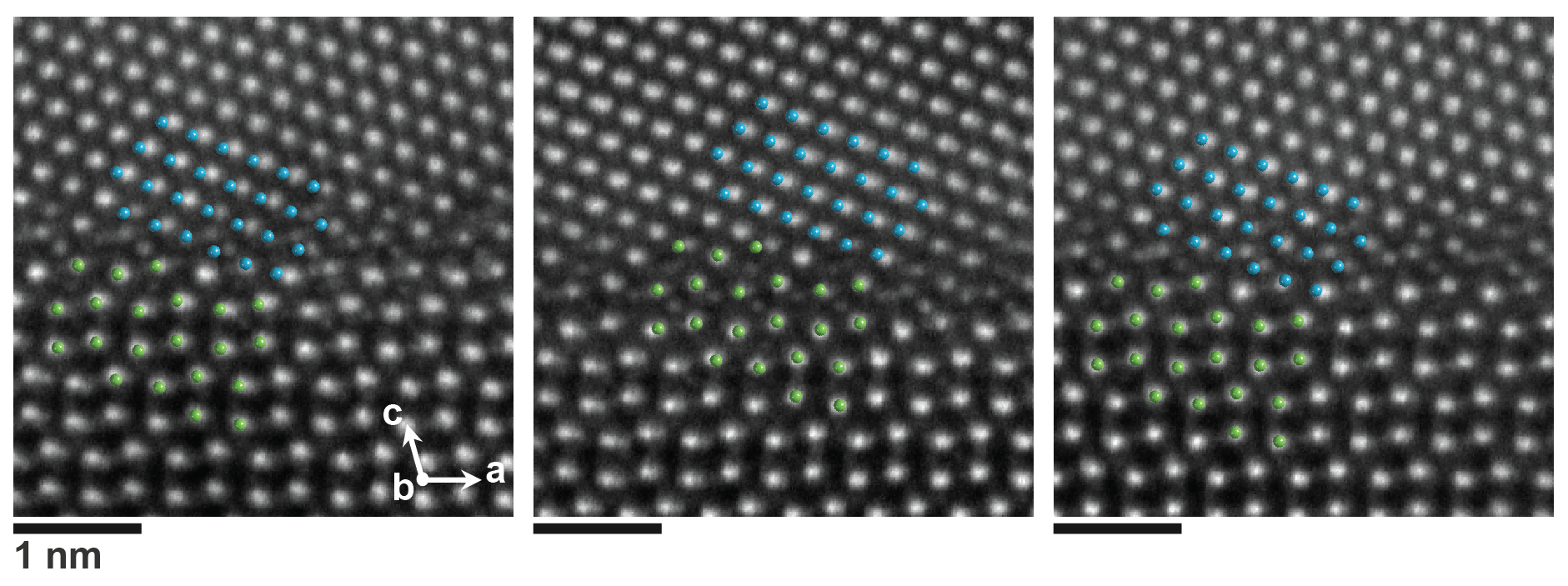}
    \caption{HAADF-STEM images of the NiO/Ga$_{2}$O$_{3}$(001) interface with the NiO/Ga$_{2}$O$_{3}$(101) model overlaid, showing variation in the alignment between the model and the atomic column contrast between images. In the model, green atoms are Ga and blue atoms are Ni. O is excluded for clarity.}
    \label{SI_101}
\end{figure*}

\begin{figure*}[t]
    \includegraphics[width=6.5in]{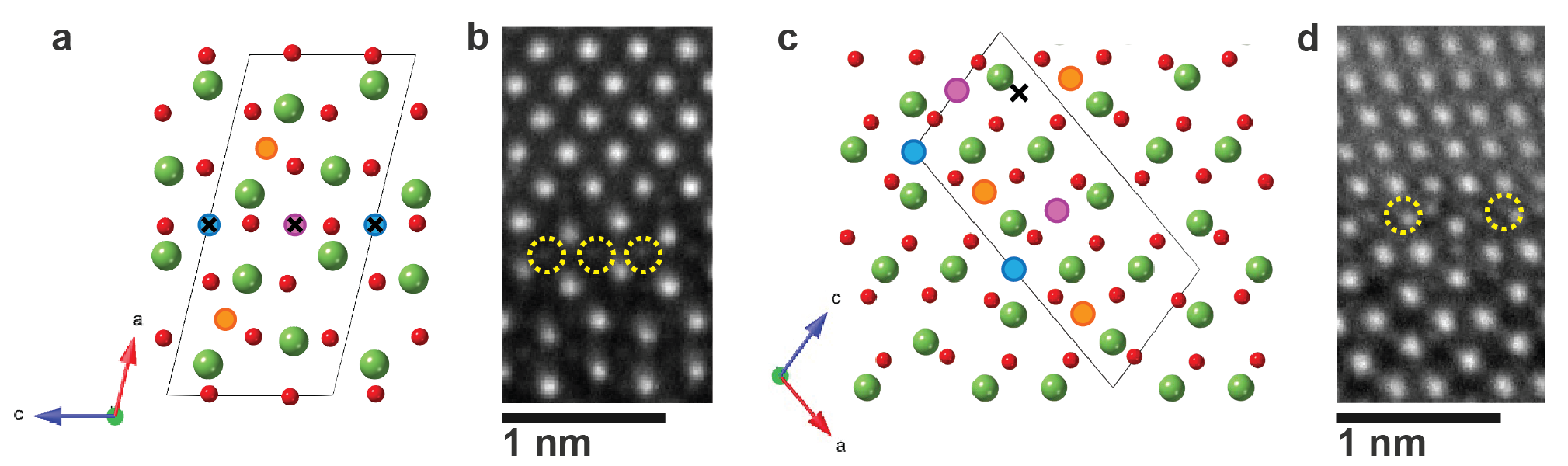}
    \caption{Ga$_{2}$O$_{3}$ interstitial sites. (a) Model of Ga$_{2}$O$_{3}$(100) viewed along the [010] direction with Ga interstitial sites indicated by pink, blue, and orange circles. (b) STEM image of NiO/Ga$_{2}$O$_{3}$(100) interface with 'extra' atomic column contrast highlighted by yellow dashed circles. (c) Model of Ga$_{2}$O$_{3}$(\={2}01) viewed along the [010] direction with the same Ga interstitial sites indicated by pink, blue, and orange circles. (d) STEM image of NiO/Ga$_{2}$O$_{3}$(\={2}01) interface with the brightest 'extra' atomic columns highlighted by yellow dashed circles. The sites of additional atomic column intensity in b and d are represented by black x's in a and c for comparison with interstitial sites.}
    \label{SI_interstitials}
\end{figure*}

Figures \ref{SI_100}-\ref{SI_001} present additional representative images for each of the NiO/Ga$_{2}$O$_{3}$ samples. Each figure shows $\sim$ 175 nm of the $\sim$ 500 nm of interface analyzed in this study for each Ga$_{2}$O$_{3}$ substrate orientation. Additional supporting figures highlighting variation in NiO(1\={1}1)/Ga\textsubscript{2}O\textsubscript{3}(101) model alignment with experimental images (Fig. \ref{SI_101}) and Ga interstitial sites in the $\beta$-Ga\textsubscript{2}O\textsubscript{3} lattice (Fig. \ref{SI_interstitials}) are included subsequently.

\section{Interface structure modeling}

We used a two-step process to generate the interface model. First, we cut the surfaces, and then we used the resulting structures to model the interface. Surfaces of $\beta$-Ga$_{2}$O$_{3}$ and NiO were generated using a bond-breaking minimization algorithm.\cite{Stevanovic_APL:2014} To define the surfaces of Ga$_{2}$O$_{3}$ slabs, both atomic coordination environment and oxygen partial pressure must be considered. Because $\beta$-Ga$_{2}$O$_{3}$ contains chemically distinct bonding environments (tetrahedrally and octahedrally coordinated Ga atoms), the energy cost of breaking these bonds differs. To account for this, the bond energy is defined in terms of chemical potential as
\begin{equation}
\varepsilon_{c,i} = \frac{\mu_c}{N_i}
\end{equation}
where $\varepsilon_{c,i}$ is the energy of one bond for the atom of component $c$ residing at site $i$, $\mu_c$ is the chemical potential of component $c$, and $N_i$ is the coordination number of that atom. 

By varying the chemical potentials within their thermodynamic stability bounds, we can model surfaces synthesized under different environmental conditions, such as O-rich or O-poor regimes. For example, under O-poor conditions, the chemical potential of oxygen is lower (more negative), which reduces the energy penalty for breaking O-related bonds and naturally leads to off-stoichiometric surface terminations. To evaluate these asymmetric, off-stoichiometric surfaces, the surrogate surface energy ($E_s^{top}$) is minimized for only the top half of the Ga$_{2}$O$_{3}$ slab:
\begin{equation}
E_s^{top} = \frac{1}{A} \sum_{i \in \{n|z_n \ge z_{mid}\}} \left[ N_i^{(slab)} - N_i^{(bulk)} \right] \varepsilon_{c,i}
\end{equation}
where $A$ is the surface area, $N_i^{(slab)}$ and $N_i^{(bulk)}$ are the coordination numbers of atom $i$ in the slab and bulk, respectively, and the summation is restricted to atoms with a $z$-coordinate greater than or equal to the middle of the slab ($z_{mid}$).

\begin{figure*}[t!]
    \centering
    \includegraphics[width=7in]{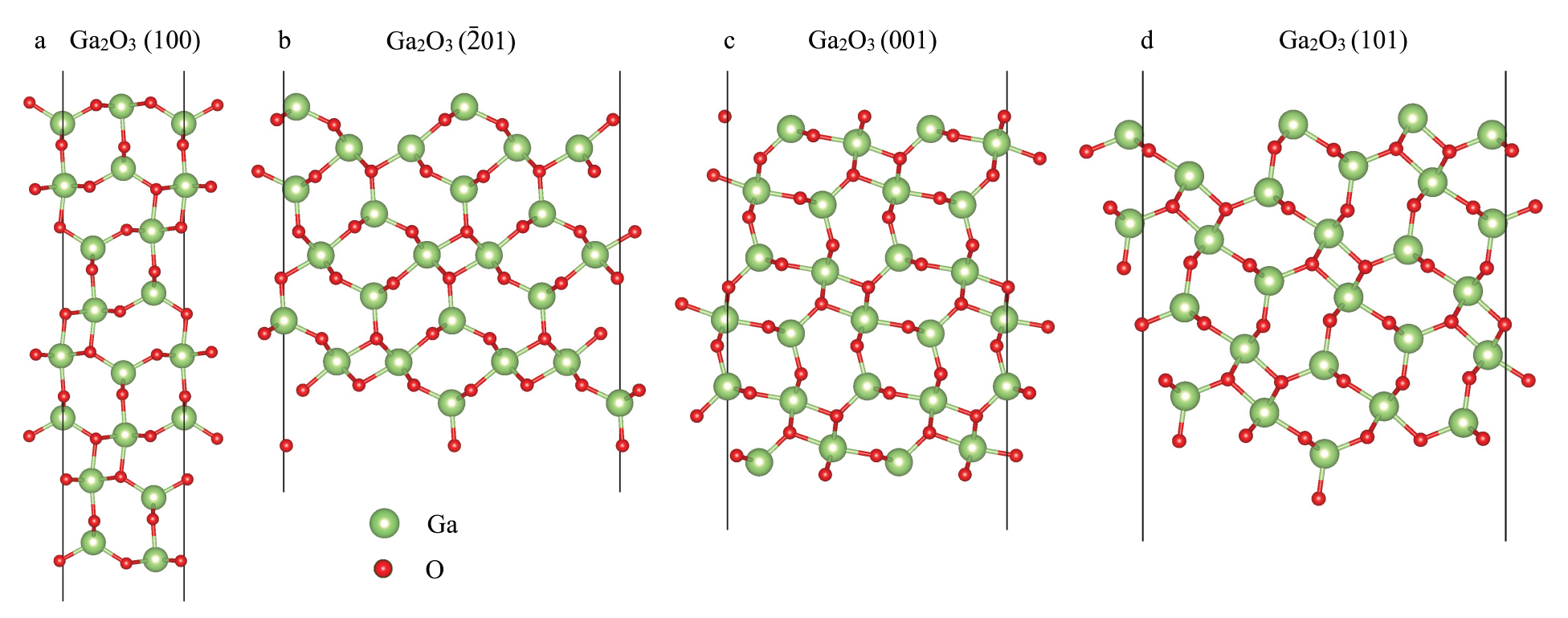}
    \caption{Different structures of the $\beta$-Ga${_2}O{_3}$ surfaces generated using the surface cutting algorithm for (a) (100), (b) ($\bar{2}$01), (c) (001), and (d) (101) facets.}
    \label{SI_Surface}
\end{figure*}

Using this approach, we systematically generated the (100), ($\bar{2}$01), (001), and (101) surfaces of $\beta$-Ga$_{2}$O$_{3}$. For NiO, we performed a search over all surface orientations within the Miller index range $-3 \leq h,k,l \leq 3$. To accurately capture the relevant thermodynamic boundary conditions, surfaces were evaluated under both O-poor and O-rich conditions using chemical potentials ($\mu$) consistent with Fitted Elemental Reference Energies (FERE) from the NLR Materials Database \cite{Stevanovic_PRB_2012}. The specific values applied are included in Table S1. 



\begin{table}[h!]
\begin{center}
\begin{tabular}{ c c c c c } 
\hline
 & \makecell{O-rich\\$\beta$-Ga$_{2}$O$_{3}$} & \makecell{O-poor\\$\beta$-Ga$_{2}$O$_{3}$ } & \makecell{O-rich\\NiO } & \makecell{O-poor\\NiO }\\
\hline
$\mu_{\text{M}}$ (eV) & -7.995 & -2.37 & -5.683 & -3.65 \\ 
$\mu_{\text{O}}$ (eV) & -4.76 & -8.51 & -5.126 & -7.159 \\ 
\hline
\end{tabular}
\caption{Chemical potential ($\mu$) values used for interface structure modeling.}
\label{tab:chem_potential}
\end{center}
\end{table}

For the $\beta$-Ga$_{2}$O$_{3}$ substrate, we selected O-rich terminations for the (100) and (001) surfaces, and O-poor terminations for the (101) and ($\bar{2}$01) surfaces, all of which are shown in Fig. \ref{SI_Surface}

For interface modeling, we adopted the structure-mapping algorithm developed by Therrien \textit{et al.} \cite{Therrien_JCP_2020,Therrien_PRAppl_2021}, in which a fixed $\beta$-Ga$_2$O$_3$ surface orientation was matched against a series of NiO surface orientations. We utilized \textit{p2ptrans}, an open-source Python package \cite{p2ptrans_2022}, to identify semi-coherent interfaces between $\beta$-Ga$_{2}$O$_{3}$ and NiO. This approach performs atom-to-atom mapping without requiring prior knowledge of the periodicity of the two materials, constrained only by a maximum allowable area strain, which we set to 8\%. Here, $\beta$-Ga$_{2}$O$_{3}$ was treated as a rigid substrate, while the NiO film was allowed to distort. 

\begin{figure*}[t!]
    \centering
    \includegraphics[width=6.5in]{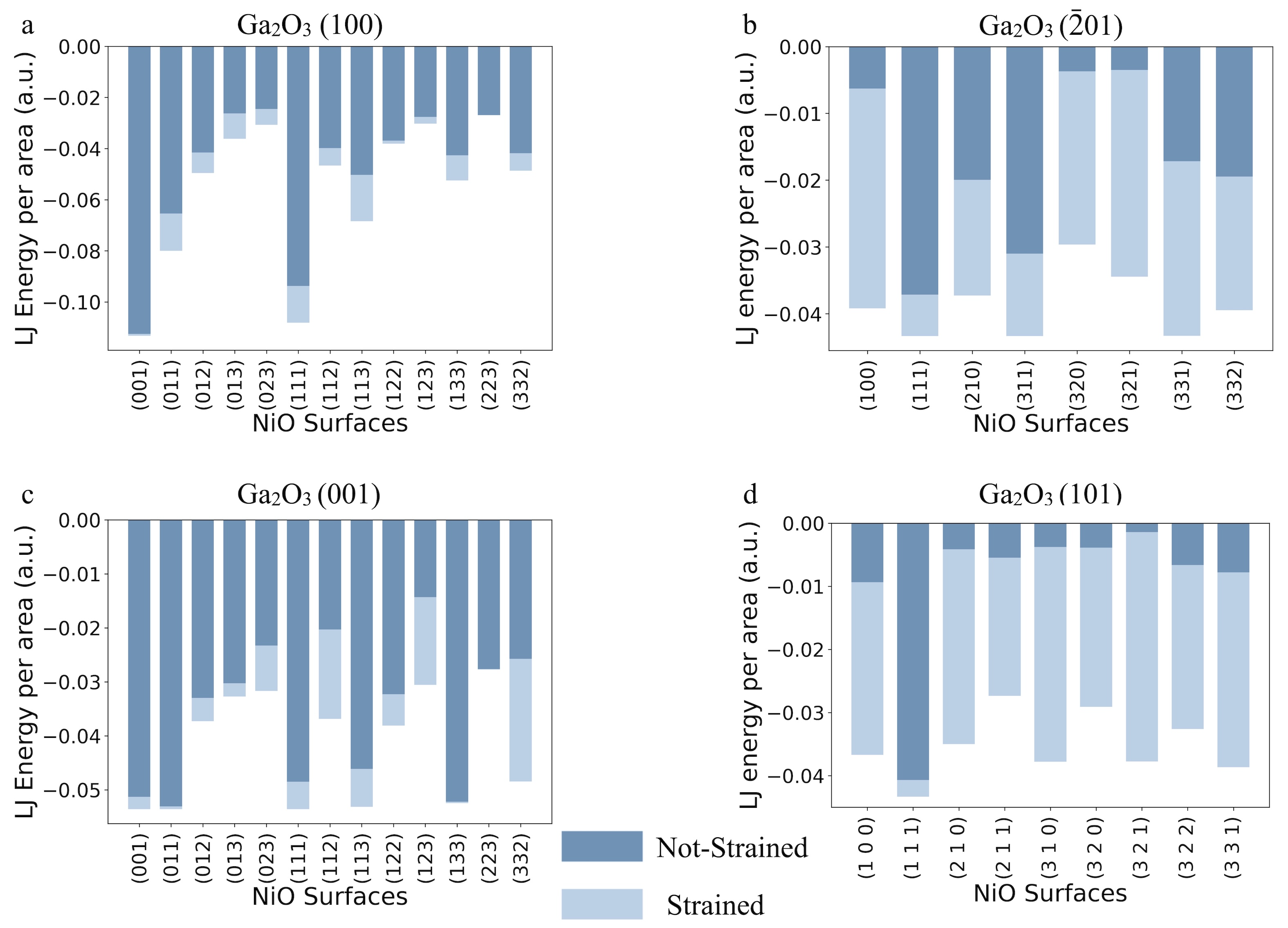}
    \caption{Structure matching results for (a) $\beta$-Ga$_{2}$O$_{3}$(100)/NiO($hkl$), (b) $\beta$-Ga$_{2}$O$_{3}$($\bar{2}$01)/NiO($hkl$), (c) $\beta$-Ga$_{2}$O$_{3}$(001)/NiO($hkl$), and (d) $\beta$-Ga$_{2}$O$_{3}$(101)/NiO($hkl$) interfaces with $h,k,l \leq 3$ and a maximum allowable area strain of 8\%. A lower Lennard-Jones (LJ) energy per unit area indicates better interface matching. Light and dark blue bars represent the goodness of match with and without strain, respectively.}
    \label{fig:interface_matching}
\end{figure*}

The goodness of match was approximated using a two-body Lennard-Jones (LJ) potential with an equilibrium radius of 2.13 \AA\ for both the Ni-O and Ga-O chemical bonds. Because cation-anion bonding is naturally more favorable than cation-cation or anion-anion interactions, interfaces involving the O-rich (100) and (001) $\beta$-Ga$_{2}$O$_{3}$ surfaces were constructed using O-poor NiO terminations and evaluated based solely on Ni-O chemical interactions. Conversely, interfaces involving the O-poor ($\bar{2}$01) and (101) $\beta$-Ga$_{2}$O$_{3}$ surfaces were paired with O-rich NiO terminations and evaluated based on Ga-O chemical interactions. We selected those specific terminations of $\beta$-Ga$_{2}$O$_{3}$ and NiO (O-poor or O-rich) based on criteria requiring both a lower absolute LJ potential energy and a minimal LJ energy difference between the strained and unstrained configurations to ensure a low strain penalty. Figure \ref{fig:interface_matching} illustrates the LJ potential as a measure of the goodness of match for various NiO surface orientations ($-3 \leq h,k,l \leq 3$), explicitly breaking down the energetic contributions with and without applied strain.

\section{Multislice STEM image simulations}

STEM images were simulated using the abTEM implementation of the multislice algorithm \cite{madsen_abtem_2021, kirkland_simulation_1987}. All simulations were calculated using a 200 kV electron probe with convergence semi-angle of 24.2 mrad and collection angles of 64 to 124 mrad, matching the experimental conditions. A defocus of 0 nm and spherical aberration coefficient of 1 $\mu$m were used. The slice thickness was 0.3 \AA. All crystal models were $\sim$5 nm thick. Following the  simulation, the output images were interpolated to increase pixel density, and a gaussian filter was applied to simulate partial coherence of the probe.

\bibliography{Ga2O3, references}